\documentclass[letters,usegraphicx,useAMS,usenatbib]{mn2e}

\usepackage{amsmath,amssymb,amsfonts,amscd}
\usepackage{aas_macros}
\usepackage{xspace}

\usepackage[T1]{fontenc}
\usepackage{ae,aecompl}
\usepackage{newtxtext,newtxmath}
\usepackage{xcolor}

\newcommand{\msol}{M_{\odot}}

\newcommand{\Msol}{M_{\odot}}

\newcommand{\zehn}[1]{10^{#1}}
\newcommand{\zehnh}[2]{{#1} \times 10^{#2}}

\newcommand{\ms}{\textrm{ms}}

\newcommand{\km}{\textrm{km}}
\newcommand{\cm}{\textrm{cm}}

\newcommand{\erg}{\textrm{erg}}

\newcommand{\sek}{\textrm{s}}
\newcommand{\isek}{\textrm{s}^{-1}}

\newcommand{\gccm}{\textrm{g\,cm}^{-3}}

\newcommand{\Gauss}{\textrm{G}}

\newcommand{\MeV}{\textrm{MeV}}

\newcommand{\modelname}[1]{\texttt{#1}}
\newcommand{\modl}[1]{model \modelname{#1}}
\newcommand{\modls}[1]{models \modelname{#1}}
\newcommand{\Modl}[1]{Model \modelname{#1}}
\newcommand{\Modls}[1]{Models \modelname{#1}}

\newcommand{\figref}[1]{Fig.\,\ref{#1}}

\newcommand{\secref}[1]{Sect.\,\ref{#1}}

\newcommand{\cf}{cf.\xspace}
\newcommand{\eg}{e.g.\xspace}

\newcommand{\nusps}{neutrinospheres\xspace}

\begin{document}

\title[Protomagnetar and black hole formation]
{Protomagnetar and black hole formation in high-mass stars}

\author[Obergaulinger \& Aloy]{
  M.~Obergaulinger$^1$,  M.\'A.~Aloy$^1$
  \\
  $^1$ Departament d{\'{}}Astronomia i Astrof{\'i}sica, Universitat de
  Val{\`e}ncia, \\ Edifici d{\'{}}Investigaci{\'o} Jeroni Munyoz, C/
  Dr.~Moliner, 50, E-46100 Burjassot (Val{\`e}ncia), Spain 
}

\maketitle

\begin{abstract}
  Using axisymmetric simulations coupling special relativistic MHD, an
  approximate post-Newtonian gravitational potential and two-moment
  neutrino transport, we show different paths for the formation of
  either protomagnetars or stellar mass black holes. The fraction of
  prototypical stellar cores which should result in collapsars depends
  on a combination of several factors, among which the structure of
  the progenitor star and the profile of specific angular momentum are
  probably the foremost. Along with the implosion of the stellar core,
  we also obtain supernova-like explosions driven by neutrino heating
  and hydrodynamic instabilities or by magneto-rotational effects in
  cores of high-mass stars. In the latter case, highly collimated,
  mildly relativistic outflows are generated. We find that after a
  rather long post-collapse phase (lasting $\gtrsim 1\,$sec) black
  holes may form in cases both of successful and failed supernova-like
  explosions. A basic trend is that cores with a specific angular
  momentum smaller than that obtained by standard, one-dimensional
  stellar evolution calculations form black holes (and eventually
  collapsars).  Complementary, protomagnetars result from stellar
  cores with the standard distribution of specific angular momentum
  obtained from prototypical stellar evolution calculations including
  magnetic torques and moderate to large mass loss rates.
\end{abstract}


\section{Introduction}
\label{Sek:Intro}

Supported by a host of observations of supernova (SN) explosions,
gamma-ray bursts (GRBs), their remnants, and the compact objects they
leave behind, current theory of stellar core collapse considers a
variety of outcomes for the post-collapse evolution of the burnt-out
cores of massive stars ($M_{\mathrm{ZAMS}} \gtrsim 8 \, \Msol $).
Plausible scenarios involve the formation of a proto-neutron star
(PNS), from whose inner core a shock wave is launched and fails to
reach the outer layers of the star but stalls inside the core.
Neutrino heating, hydrodynamic instabilities, and, possibly, other
mechanisms such as rotation and magnetic fields act against the
accretion of the surrounding shells increasing the mass of the PNS.
The balance between these effects may lead to a SN explosion in which
the accretion flows are shut down and the PNS at the centre gradually
transforms into a neutron star.  It is also conceivable that the
asymmetric geometry of the explosion runs parallel to continuing
accretion and, hence, the PNS grows beyond the maximum mass supported
against self-gravity and, finally, collapses to a black hole (BH).
The latter outcome is probably inevitable if no SN occurs at all.
Irrespective of the failure or success of a previous SN explosion,
high-energy transients are possible even after BH formation, in
particular if rapid rotation allows for a long GRB in the collapsar
model.

The first scenario (SN explosion and NS formation) seems likely
particularly for stars in the lower range of masses \citep[for recent
studies, see,
e.g.,][]{OConnor_Ott__2011__apj__BlackHoleFormationinFailingCore-CollapseSupernovae,Janka__2012__ARNPS__ExplosionMechanismsofCore-CollapseSupernovae,Ugliano_et_al__2012__apj__Progenitor-explosionConnectionandRemnantBirthMassesforNeutrino-drivenSupernovaeofIron-coreProgenitors,Sukhbold_et_al__2016__apj__Core-collapseSupernovaefrom9to120SolarMassesBasedonNeutrino-poweredExplosions,Nakamura_et_al__2015__pasj__Systematicfeaturesofaxisymmetricneutrino-drivencore-collapsesupernovamodelsinmultipleprogenitors,Bruenn_et_al__2016__apj__TheDevelopmentofExplosionsinAxisymmetricAbInitioCore-collapseSupernovaSimulationsof12-25MStars},
while we might encounter conditions favouring BH formation with or
without SN explosion among stars with higher masses.  The former
studies suggest a fairly complex dependence of the final outcome on
the progenitor conditions rather than, say, a clear threshold at a
certain mass.  We explore this complex landscape focusing on low
metallicity models, some of which have been traditionally considered
as precursors of collapsars
\citep{Woosley_Heger__2006__apj__TheProgenitorStarsofGamma-RayBursts}.
We examine in detail the post-collapse evolution of five models of
core collapse based on two $35 \, \Msol$ progenitors taken from
stellar evolution calculations (see \secref{Sek:models},
\secref{Sek:results}, and \secref{Sek:concl} for our models, results,
and conclusions, respectively).  We vary the angular momentum profile
and the magnetic field strength in the progenitor star and perform
axisymmetric simulations including special-relativistic MHD and
two-moment neutrino transport.  A similar study has been performed by
\cite{Burrows_etal__2007__ApJ__MHD-SN} and
\cite{Dessart_et_al__2008__apjl__TheProto-NeutronStarPhaseoftheCollapsarModelandtheRoutetoLong-SoftGamma-RayBurstsandHypernovae}.
However, we improve on that paper in several aspects (neutrino
transport, inclusion of special relativistic MHD and the treatment of
gravity).

\section{Models}
\label{Sek:models}

We performed axisymmetric simulations of the coupled evolution of the
stellar gas, its magnetic fields, and the neutrinos emitted by the
core solving the equations of special-relativistic
magnetohydrodynamics and neutrino transport.  To this end, we use the
Eulerian finite-volume code described in
\cite{Just_et_al__2015__mnras__Anewmultidimensionalenergy-dependenttwo-momenttransportcodeforneutrino-hydrodynamics},
which we previously had applied in core-collapse modelling
\citep{Obergaulinger_et_al__2014__mnras__Magneticfieldamplificationandmagneticallysupportedexplosionsofcollapsingnon-rotatingstellarcores}.
The most important extension w.r.t.~the latter study is the inclusion
of heavy-lepton neutrinos and the pair processes (electron-positron
annihilation and nucleonic bremsstrahlung) producing them.  We
included the effects of gravity in the neutrino transport in the
$\mathcal{O}(v)$-plus formulation of
\cite{Endeve_et_al__2012__ArXive-prints__ConservativeMomentEquationsforNeutrinoRadiationTransportwithLimitedRelativity}.

We use the equation of state (EoS) of
\cite{Lattimer_Swesty__1991__NuclearPhysicsA__LS-EOS} with an
incompressibility of $K_3 = 220 \, \MeV$ for densities above $\rho =
\zehnh{6}{7} \, \gccm$, which allows for a maximum baryonic
neutron-star mass of $M_{\mathrm{m;bry}} \approx 2.45 \, \Msol$ (we
note that all masses quoted in this article will be baryonic masses.)
At lower densities, the EoS contains electrons and positrons, photons,
and baryons.

\begin{figure}
  \centering
  \includegraphics[width=\linewidth]{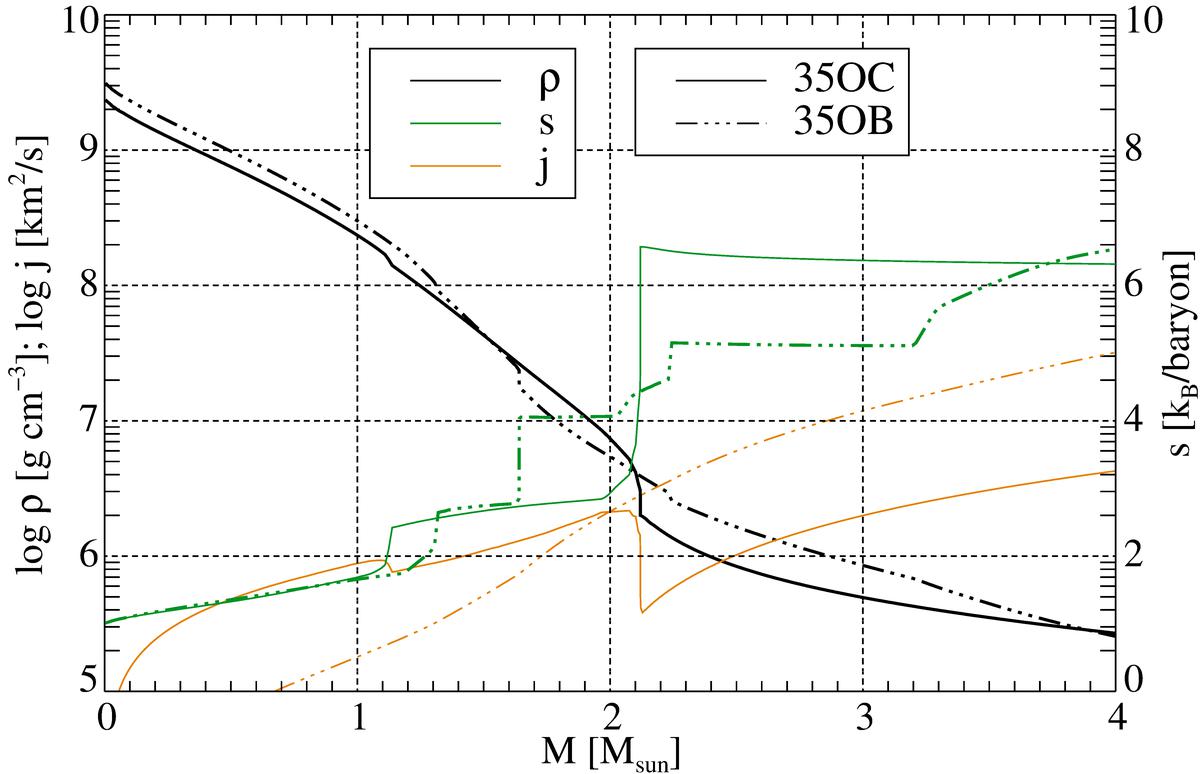}
  \caption{
    Pre-collapse profiles of mass density (black), specific
    entropy (green), and specific angular momentum (orange) of models
    \modelname{35OC} (solid lines) and
    \modelname{35OB} (dashed-triple-dotted lines) as functions of
    enclosed mass.
  }
  \label{Fig:initprof}
\end{figure}

We simulate two stars with zero-age main-sequence masses of
$M_{\mathrm{ZAMS}} = 35 \, \Msol$, one of which is considered as a
prototype progenitor of a collapsar.  The pre-collapse states of
stellar models \modelname{35OC} and \modelname{35OB} are the results
of spherical stellar-evolution computations of
\cite{Woosley_Heger__2006__apj__TheProgenitorStarsofGamma-RayBursts}
that include rotation and an approximation to a dynamo driven by
rotation and the associated magnetic transport of angular momentum
\citep{Spruit__2002__AA__Dynamo}.  The models differ only in
their mass-loss rate, whose values translate to total stellar
masses of $M_{\modelname{35OC}} = 28.1 \, \msol$ and
$M_{\modelname{35OB}} = 21.2 \, \msol$ at collapse.
The structure of the two cores differs considerably, particularly at
mass coordinates outside $2 \,\Msol$ (\figref{Fig:initprof}), i.e., in
the regions whose accretion will determine the final fate of the SN
shock wave and the compact remnant.  The progenitor star
\modelname{35OC} possesses a non-convective iron core in
(quasi-)nuclear statistical equilibrium (NSE) of
$M_{\mathrm{Fe}; \modelname{35OC}} \approx \, 2.1 \, \Msol$ surrounded
by a region of about $4 \, \Msol$ in which convection maintains a
flat entropy profile (see \figref{Fig:initprof}).  In
\modl{35OB}, the transition from the iron core to the convective
layers is made up of two distinct shells, and the jumps of entropy and
density at the first interface (at $m \approx 2.3 \, \Msol$) are less
than at the corresponding interface in \modl{35OC}, and the next, more
gradual, transition is at a rather large mass coordinate of
$m \approx 3.2 \, \msol$.  Further differences can be found in the
rotational profile: \modl{35OB} possesses a smooth,
monotonically increasing profile of specific angular momentum, $j$.
On the other hand, \modl{35OC} displays a drop of $j$ by a factor of 5
at a mass coordinate of $2.1 \, \msol$, such that the inner and outer
regions of the core have a larger and smaller $j$
than \modl{35OB} at the same mass coordinates, respectively.

We simulated versions of the former model with its original rotational
profile and either weak or very strong magnetic fields (models
\modelname{35OC-RO}, \modelname{35OC-Rw}, and \modelname{35OC-Rs},
respectively) as well as a version of the model in which we reduced
the rotational rate by a factor 4 and included a weak magnetic field
(\modl{35OC-Sw}).  To this set, we added \modl{35OB-RO}, a simulation of
core \modelname{35OB} with the original rotational profile and
magnetic field.
We initialized \modls{35OC-RO} and \modelname{35OB-RO} using the
profiles of the radial and toroidal components of the
stellar-evolution model.  These fields are fairly strong with maximum
strengths $b^{r}_{\mathrm{max}} \approx \zehnh{5}{10} \, \mathrm{G}$
and $b^{\phi}_{\mathrm{max}} \approx \zehn{12} \, \mathrm{G}$
(\modl{35OC}) and $b^{r}_{\mathrm{max}} \approx \zehnh{2}{10} \,
\mathrm{G}$ and $b^{\phi}_{\mathrm{max}} \approx \zehnh{9}{11} \,
\mathrm{G}$ (\modl{35OB}).  The artificial magnetic fields included in
\modls{35OC-Rw}, \modelname{35OC-Rs}, and \modelname{35OC-Sw} are
defined in terms of the vector potential, $\vec A$, of
\cite{Suwa_etal__2007__pasj__Magnetorotational_Collapse_of_PopIII_Stars}
\begin{equation}
  \label{Gl:b-init-p}
  ( A^{r}, A^{\theta}, A^{\phi} ) = 
  [2 (r^3 + r_0^3)]^{-1}
  \left( b_0 {r_0^3 r \sin
    \theta}, 
    0, 
    b_0 {r_0^3 r^2 \cos
    \theta}
  \right).
\end{equation}
We set the radius parameter $r_0 = \zehn{8} \, \cm$ and the
normalizations such that the poloidal and toroidal components of the
magnetic field at the origin are $b^{\mathrm{pol, tor}} = \zehn{8} \,
\mathrm{G}$ or $\zehn{10} \, \mathrm{G}$ for all models except for
\modelname{35OC-Rs}, for which $b^{\mathrm{pol, tor}} = \zehn{12} \,
\mathrm{G}$.  The exact value of the field strength in the former
class of models is not important because the field remains dynamically
insignificant in our simulations.  However, we cannot exclude that
such weak fields grow to relevant strengths by processes
\citep[magnetorotational instability, turbulent dynamos,
see][]{Moesta_et_al__2015__nat__Alarge-scaledynamoandmagnetoturbulenceinrapidlyrotatingcore-collapsesupernovae}
acting at small scales below our grid resolution.  The computational
domain extends up to $R_{\rm out}=1.4\times \zehn{10}\,$cm, spans
latitudes in the range $[0^\circ,180^\circ]$ and it is covered with
$n_r\times n_\theta=400\times 128$ zones in spherical coordinates
(logarithmically spaced in the radial direction and uniform in the
angular direction). The neutrino spectrum is discretized in 10
log-spaced bins.

\section{Results}
\label{Sek:results}

\begin{figure}
  \centering
  \includegraphics[width=\linewidth]{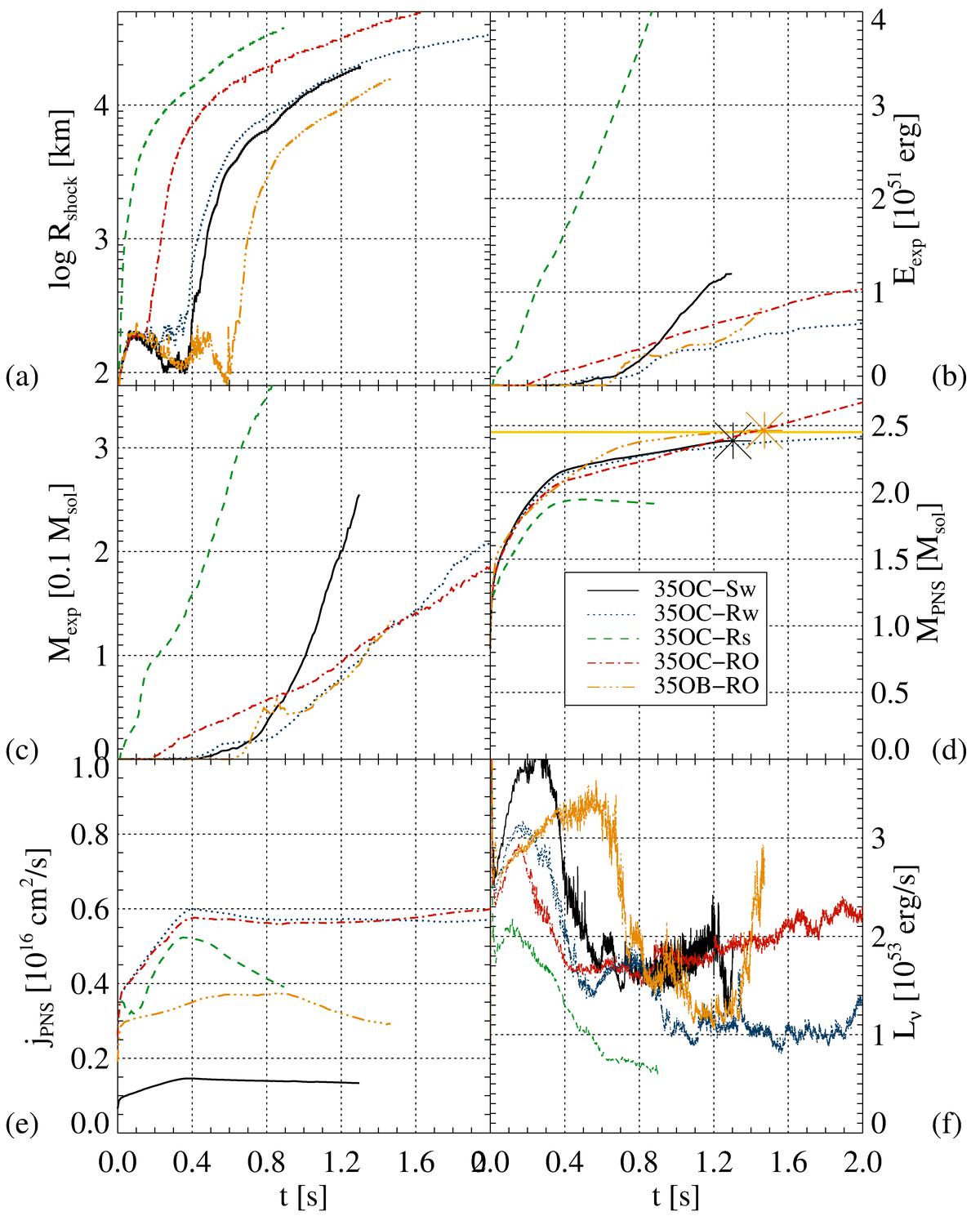}
  \caption{
    Panel \textit{(a)}: maximum shock radii of our models as a
    function of post-bounce time.
    Panels \textit{(b,c)}: total energy and mass of all unbound
    material.
    Panels \textit{(d,e)}: 
    masses and specific angular momentum of the PNS, defined as
    all mass exceeding a density threshold of $\rho_{\mathrm{PNS}} =
    \zehn{12} \, \gccm$.  The horizontal line shows the maximum mass
    of a cold-non-rotating neutron star supported by our EoS, and asterisks mark the
    instants of collapse to a BH.
    Panel \textit{(f)}: total (all flavours) neutrino
    luminosities.
  }
  \label{Fig:globals}
\end{figure}

\cite{OConnor_Ott__2011__apj__BlackHoleFormationinFailingCore-CollapseSupernovae}
demonstrated that very compact cores, characterized by a high value of
the compactness, $\xi_{m} = \frac{m/\Msol}{R_{m} / 1000 \, \km}$,
where $R_{m}$ is the radius enclosing a total mass $m$ (usually taken
to be $2.5\Msol$), tend to resist shock revival and yield failed SN
explosions that lead to BH formation by accretion onto the PNS.  The
values of $\xi_{2.5; \modelname{35OC}} = 0.49$ and $\xi_{2.5;
  \modelname{35OB}} = 0.56$ place our cores in a marginal regime where
explosions can be expected to be quite difficult to achieve.
Nevertheless, as the time evolution of the shock radii (Panel
\textit{(a)} of \figref{Fig:globals}) demonstrates, all models
eventually achieve shock revival, albeit driven by distinct
mechanisms.  The explosions are very asymmetric, thus allowing for
ongoing equatorial accretion onto the PNS, which, therefore, might
reach the maximum mass that can be supported against gravity and then
collapse to a BH.  Panels \textit{(b)} and \textit{(c)} present the
time evolution of the total (internal, kinetic, magnetic and
gravitational) energy and the mass for all the matter that is unbound,
i.e., has a positive total energy.  The former quantity is a proxy of
the explosion energy.  We find a broad range of energies and masses.
We note particularly energies in excess of $\zehn{51} \, \erg$ and
still growing when the simulations are ended.  Only \modl{35OB-RO}
shows a lower explosion energy, but we expect it to exceed $\zehn{50}
\, \erg$ if the model were run for a longer time.  In all cases, we
find ejecta masses $\gtrsim 0.2 \, \msol$, which may grow further on
longer time scales.

The high compactness of the cores \modelname{35OC} and
\modelname{35OB} translates into PNSs that are quite massive already
at their formation and grow strongly afterwards (Panel \textit{(d)} of
\figref{Fig:globals}), exceeding $2 \, \Msol$ after only 200 ms except
for \modl{35OC-Rs}, for which the PNS mass reaches a maximum of
$M_{\mathrm{PNS}} \approx 1.95 \, \msol$ at $t \approx 450 \, \ms$ and
then decreases again.  For models of core \modelname{35OC}, the
increase slows down once the interface of the iron core has been
accreted, which happens at $t \sim 300 \, \ms$.  A similar behaviour
was found in 2D; e.g.,
\cite{Cerda-Duran_et_al__2013__apjl__GravitationalWaveSignaturesinBlackHoleFormingCoreCollapse}
and in 3D models; e.g.,
\cite{Hanke_et_al__2013__apj__SASIActivityinThree-dimensionalNeutrino-hydrodynamicsSimulationsofSupernovaCores}.
\Modl{35OC-Sw} undergoes BH collapse after $\sim 1.2 \, \sec$ with a
PNS mass slightly below the EoS limit (the difference can be
attributed to our approximate GR potential).  We note that the
decrease of the mass accretion rate (the derivative of the curves in
the figure) is less pronounced for \modl{35OB-RO} than for
\modls{35OC} as the former model starts at this point to accrete the
still fairly dense intermediate layer outside of the comparably weak
density jump at $m \approx 2.3 \, \Msol$.  Hence, the PNS of
\modl{35OB-RO} exceeds the maximum mass of a cold, non-rotating NS
rather quickly.  Even though the PNS mass grows to $M_{\mathrm{PNS}}
\approx 2.6 \, \msol$, it is temporarily stabilized by thermal and
rotational support, until it collapses to a BH too.  In addition, all
of the PNSs possess fairly high rotational energy and specific angular
momentum (Panel \textit{(e)}) of the order of several $\zehn{15} \,
\cm^2 \sec^{-1}$.  The highest values are found in the rapidly
rotating versions of \modl{35OC}. \Modls{35OC-Rw} and
\modelname{35OC-RO} maintain a roughly constant level of $j \approx
\zehnh{5.7}{15} \, \cm^2 \sec^{-1}$ after $t \approx 0.4 \sec$,
whereas the decrease of the PNS mass of \modl{35OC-Rs}, setting in
around that time, is accompanied by a drop of its specific angular
momentum.  \Modl{35OC-Sw} behaves similarly to the former two models
and shows a flat specific angular momentum, though at a much lower
level.  Finally, the comparably high specific angular momentum that
\modl{35OB-RO} possesses outside a mass coordinate of $m = 2.3 \,
\msol$ (see \figref{Fig:initprof}) yields a moderate rise of $j$,
which, does not exceed that of \modls{35OC-Rw} and
\modelname{35OC-RO}, though.

The neutrino luminosities (Panel \textit{(f)} of \figref{Fig:globals})
reflect the mass accretion history with a drop corresponding to the
accretion of the interface of the iron core.  Rapidly rotating models
consistently emit less energy in neutrinos than their slowly rotating
counterparts.  Many studies of the explosion mechanism of CCSNe
\citep[\eg][]{Murphy_Burrows__2008__apj__CriteriaforCore-CollapseSupernovaExplosionsbytheNeutrinoMechanism,Nordhaus_et_al__2010__apj__Dimension_as_a_Key_to_the_Neutrino_Mechanism_of_CCSNe,Hanke__2012__apj__IsStrongSASIActivitytheKeytoSuccessfulNeutrino-drivenSupernovaExplosions,Janka__2012__ARNPS__ExplosionMechanismsofCore-CollapseSupernovae,Suwa_et_al__2016__apj__TheCriterionofSupernovaExplosionRevisited:TheMassAccretionHistory}
found a positive correlation between the neutrino luminosity of a core
and its tendency towards explosion with shock revival setting in if a
critical luminosity that depends on, among other factors, the mass
accretion rate of the core, is exceeded. The similar times of
explosion found in \modl{35OC-Rw} and \modl{35OC-Sw} seem to
contradict this connection.  An even lower luminosity can be found in
models \modelname{35OC-Rs} and \modelname{35OC-RO}, which are also the
first to explode, whereas \modl{35OB-RO} with the second lowest
luminosity of all models explodes last.  To resolve these apparent
disagreements with theoretical expectations, we have to turn to the
two-dimensional structure of the models as we shall see below in the
discussion of the individual models.

\begin{figure}
  \centering
  \includegraphics[width=1.0\linewidth]{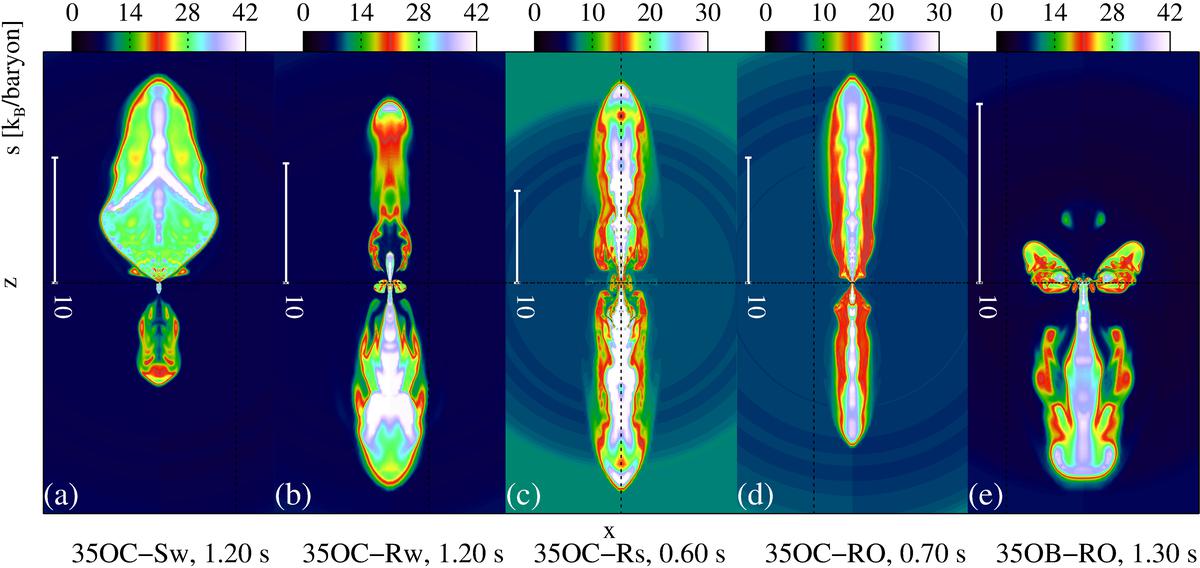}
  \caption{
    Maps of the specific entropy of the cores of all models at
    different times after core bounce as listed in the panels.  The
    length scale of the panels is represented by a ruler of length $10^4$
    km.  }
  \label{Fig:2dplots}
\end{figure}

\begin{figure}
  \centering
  \includegraphics[width=1.0\linewidth]{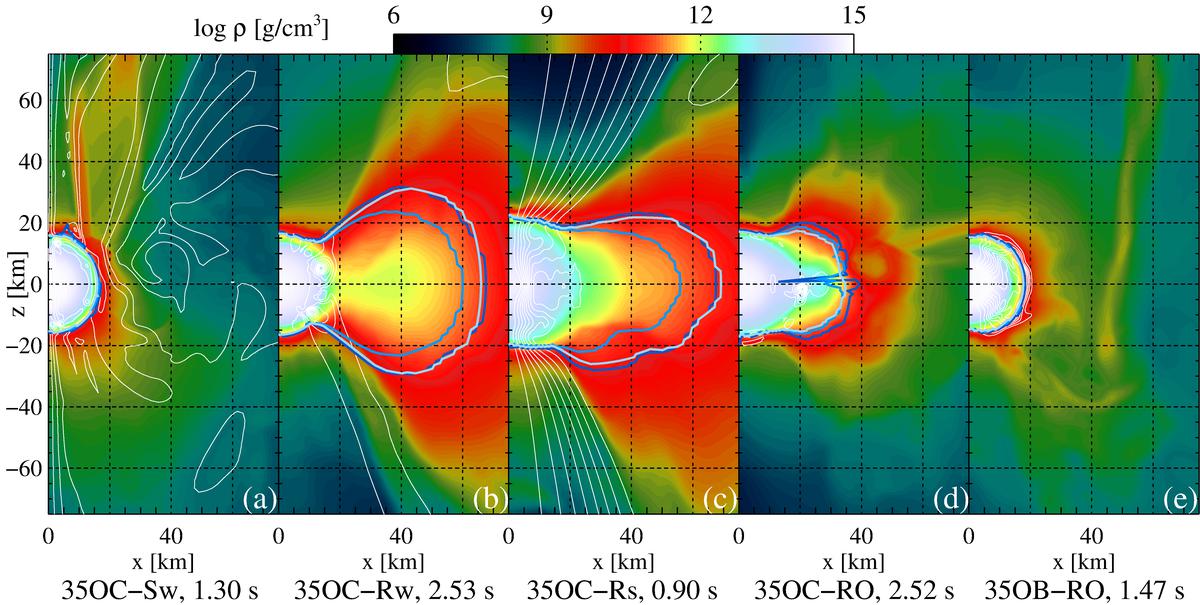}
  \caption{
    Mass density of the cores of all models prior to the
    end of the simulations (names and times listed in the
    panels).  The magnetic field is represented by field lines (white)
    and blue lines show the three \nusps.
  }
  \label{Fig:2dplots-2}
\end{figure}

The explosion of \modl{35OC-Sw} is closest to the standard paradigm of
neutrino-driven shock revival aided by hydrodynamic instabilities.
After stagnating and receding, the shock starts to expand once the
surface of the iron core has been accreted.  The process is dominated
by the highly stochastic dynamics of convection and the standing
accretion shock instability (SASI) creating large-scale bubbles of
high entropy which evolve into outflows of a degree of collimation
that varies with time, but is on average less pronounced than in other
models.  We also note that the ejecta show, due to their origin in
highly stochastic processes a strong north-south asymmetry (see Panel
\textit{(a)} of \figref{Fig:2dplots}).  Downflows remain active
between these bubbles, feeding the growth of the PNS.  Consequently,
the PNS collapses to a BH after $t \sim 1.2 \, \sec$.  We point out
that the neutrino emission is highly aspherical and the amount of
energy deposition varies strongly with time and latitude, leading to a
highly variable rate of injection of mass and energy into the northern
and southern outflows. By the time of BH formation, the core shows a
roughly spherical shape with a radius $R_{\mathrm{PNS}} \approx 17 \,
\km$ (Panel \textit{(a)} of \figref{Fig:2dplots-2}) and a small spin
parameter $a \approx 0.17$.  The high compactness of the core
translates the moderate rotational and magnetic energies into a high
surface-averaged angular velocity in excess of $\Omega_{\mathrm{PNS}}
\gtrsim 1000 \, \isek$ and surface fields of several $\zehn{14} \,
\Gauss$.

Prior to the onset of explosion, the shock surface and the
neutrino-spheres are oblate due to the strong centrifugal support in
\modl{35OC-Rw}.  Consequently, the gain layer at the pole is exposed
to an enhanced neutrino flux.  This effect locally compensates for the
reduction of the total neutrino luminosity mentioned above (\cf Panel
\textit{(f)} of \figref{Fig:globals}).  Along the poles where
rotational support is weakest, $L_{\nu}$ is even stronger than in
\mbox{\modl{35OC-Sw}}.  Thus, though residing quite deep in the
gravitational well, the polar regions of the gain layer satisfy the
conditions for an explosion as soon as the iron core interface is
accreted and the mass accretion rate and ram pressure at the shock
decrease.  We note that the latter variables are already before that
point smaller than in \modl{35OC-Sw}.  For this reason, the shock does
not retreat as much as in that model.  An explosion develops whose
geometry is similar to that of \modl{35OC-Sw} (Panel \textit{(b)} of
\figref{Fig:2dplots}).  The explosion occurs while matter is still
falling onto the PNS at low latitudes, allowing the PNS to grow.
Compared to \modl{35OC-Sw}, the time of shock revival and the
evolution of the maximum shock radius are almost unchanged, but both
their mass and energy grow slower (Panels \textit{(b,c)} of
\figref{Fig:globals}).  The degree of collimation of the outflow,
while varying with time, is higher than that of \modl{35OC-Sw}, in
particular towards the end of the simulation.  A rotationally
stabilized quasi-toroidal structure extending at the equator to
$\simeq 60$\,km (compared to a polar radius of $\simeq 16\,$km)
surrounds the inner core (see Panel \textit{(b)} of
\figref{Fig:2dplots-2}).  This equatorial torus contains a fairly
large fraction of the total angular momentum, its surface rotates
rapidly ($\Omega_{\mathrm{PNS}} > 2000 \, \isek$) and is strongly
magnetized ($b_{\mathrm{PNS}} > \zehn{14.5} \, \Gauss$).


If rapid rotation is combined with a strong magnetic field
(\modl{35OC-Rs}), a bipolar explosion develops almost immediately
after bounce (\figref{Fig:2dplots}, Panel \textit{(c)}).  Along the
poles, the shock wave never even stalls. The explosion does not result
from the reduction of the mass accretion rate because shock expansion
starts long before the surface of the iron core has fallen onto the
PNS, i.e., in the phase when $\dot{M}$ is greater than at the time of
shock expansion of models \modelname{35OC-Sw} and \modelname{35OC-Rw}.
Later on, mass accretion, occurring mainly via equatorial downflows is
weaker and is after $t \approx 400 \, \ms$ even entirely quenched by
the expanding ejecta.  Consequently, the PNS mass is by the end of the
simulation $\simeq 0.2 \, \Msol$ smaller than in the other two models,
and we do not expect the PNS to collapse to a BH for many dynamical
time scales.  The explosion is not caused mainly by neutrinos, whose
luminosity is weaker than in any other model.  Instead, the ejecta are
generated magneto-rotationally.  By the end of the simulation, the
outflows have reached velocities of $v_r \sim c/3$, and the speed of
the jet head is close to $c/5$.  The oblate geometry of the PNS is
even more pronounced than in the models discussed so far (Panel
\textit{(c)} of \figref{Fig:2dplots-2}).  The surface is at
$\Omega_{\mathrm{PNS}} \approx 1000 \, \isek$ and $b_{\mathrm{PNS}}
\lesssim \zehn{15} \, \Gauss$, i.e., \modl{35OC-Rs} rotates slowlier
and is more magnetized than \modl{35OC-Rw}.  The axisymmetry of our
models restricts them compared to three-dimensional models
\citep{Mosta_et_al__2014__apjl__MagnetorotationalCore-collapseSupernovaeinThreeDimensions}
since first, they do not observe the quenching of the accretion flow
by the ejecta and, second, their jets exhibit 3d instabilities absent
in our models.

The behaviour of \modl{35OC-RO} (panel \textit{(d)} of
\figref{Fig:2dplots}) lies in between those of \modls{35OC-Rw} and
\modl{35OC-Rs}.  Shock revival occurs at $t \approx 200 \, \ms$ due to
neutrino heating concentrated at the poles.  Strong magnetic fields
partially suppress lateral motions in the gain layer.  Consequently,
fluid elements are not as likely to leave the region of maximum
heating rate sideways towards the equator as in \modl{35OC-Rw}.  The
explosion caused by neutrino energy deposition aided by magnetic
fields shares many characteristics with standard neutrino-driven
explosions such as a certain degree of stochasticity that manifests
itself in north-south asymmetry of the ejecta.  The strong magnetic
field transports angular momentum from the interior of the PNS to its
outer layers, which as a consequence possess an excess of rotational
support and assume an oblate form (Panel \textit{(d)} of
\figref{Fig:2dplots-2}).  At the termination of the simulation, the
PNS is very massive ($M_{\mathrm{PNS}} \gtrsim 2.6 \, \msol$), but not
very compact (the polar and equatorial radii are between 16 and 40 km)
due to its fast rotation (the effective spin parameter is $a \approx
0.6$), and it does not seem to be at the verge of forming a BH.  Its
surface rotation and magnetic field are characterized by high values
of $\Omega_{\mathrm{PNS}}\lesssim 2000 \, \isek$ and $b_{\mathrm{PNS}}
\gtrsim \zehnh{5}{14} \, \Gauss$.

With a magnetic field too weak for a magneto-centrifugal explosion,
\modl{35OB-RO} explodes similarly to \modl{35OC-Rw}, albeit much later
($t \sim 600 \, \sek$).  We refer to Panel \textit{(e)} of
\figref{Fig:2dplots} showing the morphology of the ejecta at late
times.  The delay is caused by the high mass accretion rates and ram
pressures at the shock persisting until the accretion of the interface
at $m = 2.3 \, \msol$ (\cf \figref{Fig:initprof}).  Because at most
locations in the core the initial density is higher than in
\modl{35OC}, the PNS mass grows rapidly and reaches $M_{\mathrm{PNS}}
= 2.45 \, \msol$ already at $t \approx 1.2 \, \sek$.  At that point,
the PNS (at a spin parameter $a \approx 0.45$) is rather compact and
moderately oblate with radii between 16 (poles) and 22 km (Panel
\textit{(e)} of \figref{Fig:2dplots-2}).  It rotates at
$\Omega_{\mathrm{PNS}} > 3000 \, \ms$ and possesses a field strength
of $b_{\mathrm{PNS}} \sim \zehnh{6}{14} \, \Gauss$.  The model forms a
moderately rotating BH after about 1.4 s.

\section{Discusion and conclusions}
\label{Sek:concl}

To assess their viability as candidates for supernova explosions and
as progenitors of long GRBs, we studied the collapse and the
post-bounce evolution of the cores of stars of initial masses of
$M_{\mathrm{ZAMS}} = 35 \, \Msol$ with different degrees of rotation
and magnetic fields in neutrino-MHD simulations.  Our five simulations
show very distinct dynamics: (1) explosions within the standard
neutrino-driven paradigm of supernova theory, but followed by collapse
to a BH; (2) rapid rotation concentrating neutrino emission along the
rotational axis and generating bipolar explosion; (3) early
magneto-rotational explosions launching moderately relativistic
outflows.

All explosions have a very prolate morphology.  The energies and
masses of the ejecta differ strongly among models, with the most
violent explosion, viz.~one launched by the magneto-rotational
mechanism, exceeding an energy of
$E_{\mathrm{exp}} = \zehnh{4}{51} \, \erg$ and a mass of
$M_{\mathrm{exp}} = 0.38 \, \msol$ at the end of the simulation, while
the others achieve energies of the order of
$E_{\mathrm{exp}} = \zehn{51} \, \erg$ that are still growing when we
had to terminate the computation.

In most cores, accretion onto the PNS goes on after the onset of
explosion, thus making a later collapse to a BH quite likely, and in
two models BH collapse was actually found. In such a case, we expect
that the models with the most rapid rotation might produce a second
energy release in which hyperaccretion onto the newly formed BH drives
another generation of outflows along the rotational axis into the
low-density funnel left behind by the initial outflows.
\cite{Dessart_et_al__2008__apjl__TheProto-NeutronStarPhaseoftheCollapsarModelandtheRoutetoLong-SoftGamma-RayBurstsandHypernovae}
found that model \modelname{35OC} is very susceptible to early
magneto-rotational explosions that inhibit the growth of the PNS mass
and make a later collapse to a BH unlikely, thus diminishing the
prospects of a collapsar-type progenitor. Indeed, the reduction of the
rotation rate (\modl{35OC-Sw}) as well as a different profile of the
core (\modl{35OB-RO}) allow for BH formation and, thus, an evolution
within the collapsar scenario. Our strongly magnetized models suggest
similar conclusions, but indicate a greater possibility for GRBs
powered by a PM central engine
(\citealt{Metzger_et_al__2011__mnras__Theprotomagnetarmodelforgamma-raybursts};
M11 hereafter).  Even though simulations on longer times are required
to settle this issue, we already observe that the way towards the
fiducial conditions assumed in M11 is more complex than sketched by
the former authors. The main reasons for the discrepancy arise from
two facts. First, the equatorial accretion along with bipolar mass
loss, both of variable strength, are active throughout the post-bounce
time here presented.  Consequently, the mass of the PNS may change
either increasing (\modelname{35OC-RO}) or slightly decreasing
(\modelname{35OC-Rs}).  It should be noted that without the
restriction of axisymmetry, the dynamics of the accretion flows
feeding the PNS might be considerably more complex and BH formation
might be less likely.  Second, the PNS shape of potential PMs is very
oblate (i.e., not spherical).  Some of the key parameters in the PM
model are the period ($P_0$), the radius and the mass of the PNS at
the moment of {\em birth} of the PM (which can be taken $\sim 10\,
\sek$ post bounce when the contraction of the PNS practically ceases
in the M11 model), as well as the poloidal component of the magnetic
field on the surface of the PNS.  The time evolution of all of them
during the first 2\,s post bounce does not follow the simple paths
sketched in M11. We estimate for models \modelname{35OC-RO} and
\modelname{35OC-Rs} values $P_0\simeq 0.1-0.25\, \ms$ extrapolating
the evolution of the PNS after 50\,ms post bounce. These values are
smaller than the ones computed for stable and isolated neutron stars
($P_0\gtrsim 1\, \ms$) of smaller masses than the PNSs we consider
here (e.g., \citealt{Strobel:1999} make their estimates for a mass of
$1.35\,M_\odot$). Likewise, M11 assume that the magnetic field energy
is a fraction $\epsilon_B\sim 10^{-3}$ of the gravitational energy of
the PNS. Our potential PM models show $\epsilon_B\gtrsim
10^{-2}$. However, the poloidal field at the PNS surface is smaller
than assumed in M11. For \modl{35OC-Rs}, $b_{\mathrm{PNS}}^{\rm pol}
\simeq b_{\mathrm{PNS}}^{\rm tor} \sim \zehnh{5}{14} \, \Gauss$, while
for \modl{35OC-RO} $b_{\mathrm{PNS}}^{\rm pol} \simeq 0.1
b_{\mathrm{PNS}}^{\rm tor} \sim \zehnh{5}{13} \, \Gauss$. Finally, we
note that potential PM models display an early phase of magnetic field
amplification where $b_{\mathrm{PNS}}$ quickly ($t\simeq 0.4\,$sec)
settles to a value $\sim \zehnh{5}{14} \, \Gauss$. Models evolving
into collapsars need $t\simeq 1\,$sec to arrive to similar values of
$b_{\mathrm{PNS}}$, and once at that level, they display variations in
strength by factors $\sim 3$ within timescales of 100\,ms.

We close with a few comments on the shortcomings and uncertainties of
our current study.  While general-relativistic versions of our models
might produce slightly different results, even shift the border
between SN explosion and BH collapse somewhat, we think that the
pseudo-relativistic potential we chose is a rather accurate
approximation and affects only the criteria for, but not the existence
of the scenarios summarized above.  In this context, we mention in
particular that our approach does not include energy losses due to
gravitational waves.  However, the rotational frequencies required for
this effect to play a role exceed those of our models, and hence we
deem its neglect insignificant.  Our treatment of neutrino-matter
interactions can be improved by using more recent opacity models, but
tentatively expect that the uncertainty coming from that source is
also more in the realm of minor quantitative shifts in the parameter
space rather than a complete change of our conclusions.  The most
important limitation is axisymmetry, as three-dimensional effects
might considerably modify the topology of the flows
\citep{Winteler_et_al__2012__apjl__MagnetorotationallyDrivenSupernovaeastheOriginofEarlyGalaxyr-processElements,Mosta_et_al__2014__apjl__MagnetorotationalCore-collapseSupernovaeinThreeDimensions}.
While very interesting, three-dimensional long-term simulations of
several models are still very challenging computationally and, thus, a
topic for future research.

\section{Acknowledgements}
\label{Sek:Ackno}
We thank O.~Just and Th.~Janka for valuable help.  We
acknowledge the support from the European Research Council (grant
CAMAP-259276), and the partial support of grants AYA2015-66899-C2-1-P,
AYA2013-40979-P and PROMETEO-II-2014-069 and computational time on the
clusters \textit{Pirineus}, \textit{Picasso}, \textit{MareNostrum},
and \textit{Tirant} under grants AECT-2016-1-0008, AECT-2016-2-0012,
and AECT-2016-3-0005 of the Spanish Supercomputing Network .

\end{document}